\documentclass[11pt]{article}
\usepackage{epsf}
\title{\bf Inclusive cross-sections for gluon production
in collision of two projectiles on two targets in the BFKL approach}
\author{M.A.Braun\\
Dep. of High Energy physics,
 Saint-Petersburg State University,\\
198504 S.Petersburg, Russia}

\newcommand\lra{\leftrightarrow}
\newcommand\beq{\begin{equation}}
\newcommand\eeq{\end{equation}}

\def\tf{\tilde{f}}

\begin{document}

\maketitle

{\bf Abstract}\\
Inclusive cross-sections for gluon production in collision of two
mucleons with two nucleons are studied in the BFKL approach.
Various contributions include emission from the pomerons attached
to the participants, from the BFKL interactions in between these
pomerons and from the intermediate BKP state. The last contribution
may be observable provided the growth with energy of the pomeron
contribution is tamed in accordance with unitarity.

\section{Introduction}
In our previous paper ~\cite{braun1} we have derived the forward
scattering amplitude for two-projectles-two target collisions
at high energies in the BFKL approach.
Its immediate application is to the cross-section for deuteron-deuteron
collisions, although it also concerns a part of
heavy-nucleus-heavy-nucleus collisions due to interaction of two
pairs of nucleons. A remarkable result found in ~\cite{braun1}
is that the cross-section contains a contribution from the intermediate
state consisting of 4 reggeized gluons in the octet colour state between
the neghbours (BKP state ~\cite{bartels,kwie}). In this paper we study the inclusive
cross-section for gluon production for the same process.
Our main result is that, similar to the total cross-section, the
intermediate BKP state gives a non-zero contribution. This should be
compared to the (approximate) expression derived in the JIMWLK approach
~\cite{dusling} in which such contribution is absent and only appears for
the double inclusive cross-sections ~\cite{double}.

\begin{figure}[h]
\leavevmode \centering{\epsfysize=0.3\textheight\epsfbox{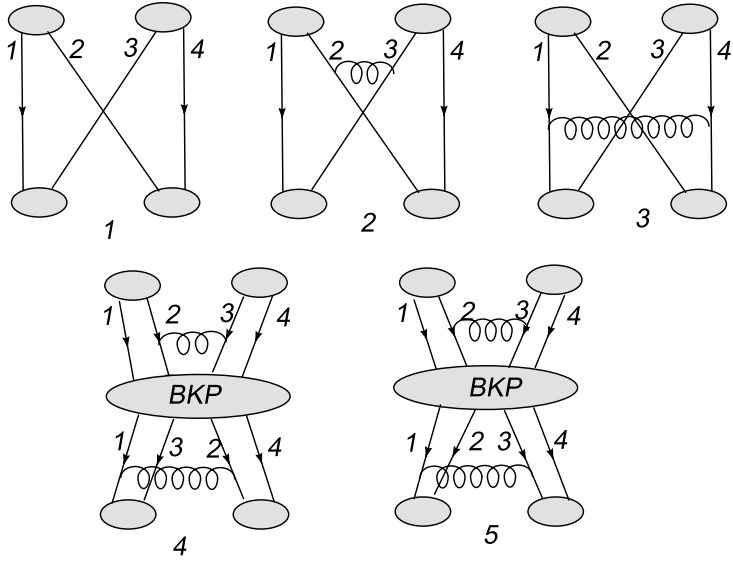}}
\caption{ The scattering ampitude for the collision of two projectiles
on two targets}
\label{fig1}
\end{figure}

As derived in ~\cite{braun1} in the lowest order in $\alpha_sN_c$,
assumed small, the scattering ampitude for the collision of two projectiles
on two targets is described by diagrams shown in Fig. \ref{fig1}.
Diagram 1  corresponds to direct sewing of the four pomerons attached
to the projectile and target with redistribution of colour. Diagrams
2 and 3 describe the situation when the gluons connecting
the projectiles and targets once interact between themselves.
Diagrams 4 and 5 cover all the rest cases when the exchanged gluons
interact at least twice. The state formed between these interactions is the
BKP state made of 4 reggeized gluons. The observed gluon may be emitted either
from the pomerons themselves, or from the interactions
between the exchanged gluons, both explicitlly appearing in the
diagrams of Fig. \ref{fig1} and implicit in the BKP state.
Correspondingly we shall study all contributions successively:
the contributions from the pomerons in Section 2, from the explicitly shown
interactions in Sections 3,4 and 5 and from the BKP state in Section 6. Section
7 is devoted to some conclusions.

The high-energy part $H$ of the forward scattering amplitude in nucleus-nucleus
scattering can be presented in the form
\beq
H=-i(2\pi)^2\delta(\kappa_+)\delta(q_-)N_c^2(kl)^2 D.
\label{defh}
\eeq
Here $\kappa$ and $q$ are the momenta transferred to the projectile nucleus
and target
nucleus respectively with $\kappa_-=\kappa_\perp=q_+=q_\perp=0$.
Factor $N_c^2(kl)^2$ is present in all diagrams, so that it is
convenient to separate it. $D$ stands for 'diagram' and gives the
contribution from the diagram for the $S$-matrix
(hence $-i$ in (\ref{defh}).
At overall rapidity $Y$ and  fixed impact papameter $b$ the
total cross-section for nucleus-nucleus scattering
corresponding to two-nucleon-two-nucleon
interaction  is related to function $D$ as (see appendix in ~\cite{braun1})
\beq
\sigma_{AB}(Y,b)=\frac{1}{4}
A(A-1)B(B-1)T^{(2)}_{AB}(b) D(Y),
\eeq
where the transverse density for two pairs of participants is
\beq
T^{(2)}_{AB}(b)=\int d^2b_Ad^2b_BT^2_A(b_A)T^2_B(b_B) \delta^2(b_A-b_B-b)
\eeq
For deuteron-deuteron scattering one finds instead
\beq
\sigma_{dd}(Y)=\frac{1}{4}\Big<\frac{1}{2\pi r^2}\Big>_A
\Big<\frac{1}{2\pi r^2}\Big>_B D(Y).
\eeq
For the inclusive cross-section the relevant high-energy part
can be presented in the same manner
\beq
H(Y,y,k)=-i(2\pi)^2\delta(\kappa_+)\delta(q_-)N_c^2(kl)^2 F(Y,y,k),
\label{defh1}
\eeq
where $y$ and $k$ are the rapidity and transverse momentum of
the observed gluon. The inclusive cross-section for nucleus-nucleus
scattering  is then
\beq
I_{AB}(Y,y,k,b)\equiv\frac{(2\pi)^2d\sigma_{AB}}{d^2kdyd^2b}=\frac{1}{4}
A(A-1)B(B-1)T^{(2)}_{AB}(b) F(Y,y,k)
\label{inclaa}
\eeq
and for deuteron-deuteron scattering
\beq
I_{dd}(Y,y,k)=\frac{1}{4}\Big<\frac{1}{2\pi r^2}\Big>_A
\Big<\frac{1}{2\pi r^2}\Big>_B F(Y,y,k).
\label{incldd}
\eeq
In the following we sometimes suppress the common arguments $Y,y,k$ in
our formulas.

After separation of the $\delta$-functions
in Eq. (\ref{defh}) or Eq. (\ref{defh1})  diagrams  contain internal
longitidinal integrations over intermediate gluon momenta. For the observed
intermediate
gluon these integrations are included in the definition of $I$
leaving only factor $1/4\pi$. Other interactions may refer to either
real gluons in the intermediate state or virtual gluons inside the
production amplitudes. If the gluon is real longitudinal integrations
reduce to the integration over its rapidity with the same
factor $1/4\pi$. If the gluon is virtual integration of
its propagator lifts one of the integrations leaving
the integration over its rapidity with an additional factor
$-i/4\pi$ (see ~\cite{braun1}). Taking into account that
inclusion of the virtual gluon provides an additional factor $(-i)^3$
we find that in the end
virtual integrations give the same result as real ones. This is important for
our calculations: the contribution from the internal gluon line does not depend
on whether it refers to the real gluon (is 'cut') or the virtual one
(is 'uncut'). Apart from integrations over the unobserved or virtual
gluon rapidities, functions $D$ and $F$ ar just contribution from the
diagrams with transverse integrations over the gluon momenta in accordance
with the conservation laws.

\section{Contribution from the pomerons}
\begin{figure}[h]
\leavevmode \centering{\epsfysize=0.1\textheight\epsfbox{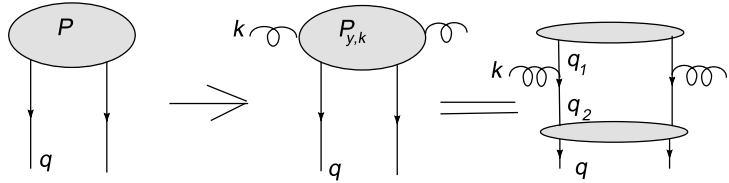}}
\caption{'Opening' of the pomeron }
\label{fig2}
\end{figure}

For the following note that in our kinematics the pomeron wave function
is real.
The inclusive cross-section for the production of a gluon with rapidity
$y$ and transverse momentum $k$ from the pomeron is well-known. It
corresponds to substitution of the pomeron $P(Y-y',q)$ by the inclusive
cross-section $P_{y,k}(Y,y',q)$ obtained by 'opening' one of the BFKL
interactions inside as shown in Fig. \ref{fig2}.

Explicitly
\beq
P_{y,k}(Y,y',q)=\int\frac{d^2q_1d^2q_2}{(2\pi)^2}\delta^2(q_1-q_2-k)
P(Y-y,q_1)v(q_2,-q_2|q_1,-q_1)G_(y-y',q,q_2),
\label{ip}
\eeq
where $v$ is the forward BFKL interaction in
the form symmetric respective to the
initial and final states
\beq
v(q,-q|k,-k)=4\alpha_sN_c\frac{1}{(q-k)^2}.
\eeq

The total inclusive cross-section from the pomerons in the diagrams
of Fig. \ref{fig1} will be obtained if we cut the diagrams
in all possible ways and substitute one of the cut pomerons according to
Eq. (\ref{ip}). A particular cut may also pass or not pass through
explicit interactions in Fig. \ref{fig1}, 2-5 or inside the BKP state.
The resulting cross-section from the pomerons will not depend on whether
these extra interactions are cut or not, since the cut BFKL interaction is
equal to uncut ones. So one can study different contributions from the
pomerons forgetting about these extra interactions, that is just from
Fig. \ref{fig1},1.

\begin{figure}[h]
\leavevmode \centering{\epsfysize=0.3\textheight\epsfbox{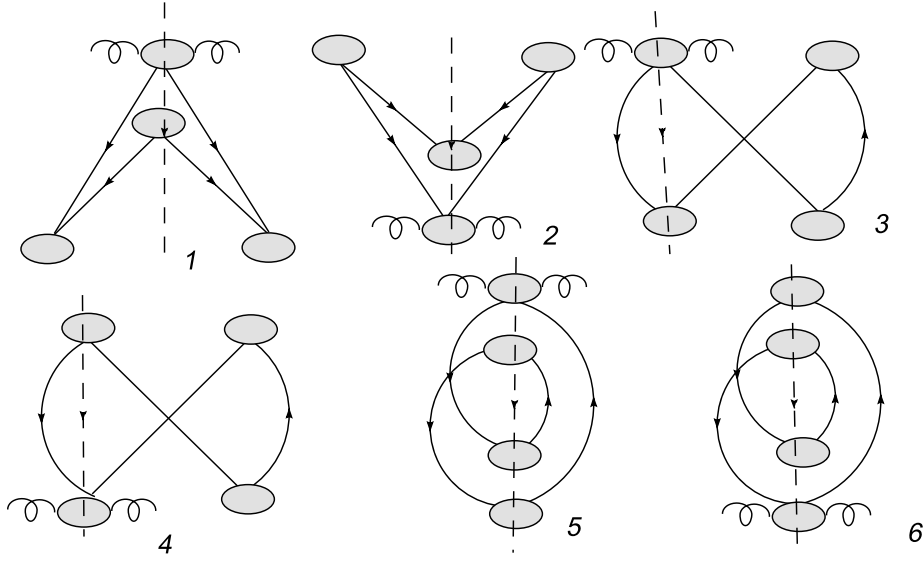}}
\caption{Different cuts in the amplitude shown in Fig. \ref{fig1},1 }
\label{fig3}
\end{figure}

The forward amplitude corresponding to Fig. \ref{fig1},1 derived in
~\cite{braun1} is given by
\beq
D_1=-\frac{\partial}{\partial Y}
\int_0^Ydy'\int \frac{d^2q}{(2\pi)^2}P^2(Y-y',q)P^2(y',q).
\label{fina1}
\eeq
Contributions to the inclusive cross-section will be given
by different cuts in the amplitude as a whole. These cuts are
shown in Fig. \ref{fig3}.
Diagrams 1 and 2 correspond to diffractive
configurations respective to the target (DT) or projectile (DP). Diagrams 3
and 4 illustrate the single cut configuration (S) in which one projectile
and one target are cut. Diagrams 5 and 6 show the double cut configuration
(DC) in which both projectiles and both targets are cut.

We start with the diffractive contributions.
The one respective to the target (Fig. \ref{fig3},1) gives the contribution
to the high-energy part $F$
\beq
F_{DT}^P=4\frac{\partial}{\partial Y}
\int_0^Ydy'\int \frac{d^2q}{(2\pi)^2}P_{y,k}(Y,y',q)P^(Y-y',q)P^2(y',q).
\label{fpdt}
\eeq
Coefficient 4 takes into account two projectiles and two targets.
The diffractive contribution respective to the projectile
(Fig. \ref{fig3},2) gives
\beq
F_{DP}^P=4\frac{\partial}{\partial Y}
\int_0^Ydy'\int \frac{d^2q}{(2\pi)^2}P^2(Y-y',q)P_{y,k}(y',0,q)P(y',q).
\label{fpdp}
\eeq

The single cut contributions (Figs. \ref{fig3},3 and 4)
enter with the minus sign. In fact they contain
one exchanged reggeon on the left and three on the right giving
$i(-i)^3=-1$.
Their contribution is
\beq
F_{S}^P=-4\frac{\partial}{\partial Y}
\int_0^Ydy'\int \frac{d^2q}{(2\pi)^2}\Big(
P_{y,k}(Y,y',q)P(Y-y',q)P^2(y',q)+
P^2(Y-y',q)P_{y,k}(y',0,q)P(y',q)\Big).
\label{fps}
\eeq
Coefficient 4 again takes into account interchanges
of the two projectiles and two targets.

Finally  the DC contribution (Figs. \ref{fig3},5 and 6) gives
\beq
F_{DC}^P=4\frac{\partial}{\partial Y}
\int_0^Ydy'\int \frac{d^2q}{(2\pi)^2}\Big(
P_{y,k}(Y,y',q)P(Y-y',q)P^2(y',q)+
P(Y-y',q)P_{y,k}(y',0,q)P(y',q)\Big).
\label{fpdc}
\eeq
Coefficient 4 takes into account two projectiles and two targets
and two different diagrams for the DC configuration.

In the sum  the S and DC contributions cancel and we
are left with only the diffractive contributions, which gives the total
$F$ from the gluons inside the pomerons
\beq
F_{1}^P=F_{DT}^P+F_{DP}^P.
\label{fp}
\eeq

As mentioned, inclusion of other interactions in between does not change
the form of the result, which is obtained from the  formulas
for the forward amplitude making the substitutions (\ref{ip}) in the
same manner as above.
The amplitude corresponding to Figs. \ref{fig1},2+3 is ~\cite{braun1}
\beq
D_{2+3}=2\int_0^Ydy'\int\frac{d^2qd^2q'}{(2\pi)^4}
<q,q'|H|q',q>
P(Y-y',q)P(Y-y',q')P(y',q)P(y',q'),
\label{fina2}
\eeq
where $H$ is the full infrared stable BFKL interaction. For gluons 1 and 2
\beq
H_{12}=-\omega_1-\omega_2-v_{12},
\eeq
where $\omega_{1,2}$ are the gluon Regge trajectories.
The high-energy function $F$ corresonding to these diagrams will be
given by
\[
F_{2+3}^P=8\int_0^Ydy'\int\frac{d^2qd^2q'}{(2\pi)^4}<q,q'|H|q',q>
\Big(P_{y,k}(Y,y',q)P(Y-y',q')P(y',q)P(y',q')\]\beq+
P(Y-y',q)P(Y-y',q')P_{y,k}(y',0,q)P(y',q')\Big).
\label{fp23}
\eeq

For the diagram in Fig. \ref{fig1},4 with the BKP state
we have the amplitude
\[
D_{3a}=\int_0^Y dy'\int_0^{y'}dy''
\int\prod_{j=1}^4\frac{d^2q'_j}{(2\pi)^2}\,\frac{d^2q_j}{(2\pi)^2}
2\pi\delta^2\Big(\sum_{j=1}^4q'_j-\sum_{j=1}q_j\Big)\]\beq
P(Y-y',q_1)P(Y-y',q_4)
<q_1,-q_1,-q_4,q_4|M^{(a)}(y'-y'')|q'_1,-q'_4,-q'_2,q'_4>
P(y'',q'_1)P(y'',q'_4).
\label{a3ta}
\eeq
Here the operator $M^{(a)}$ acting  in the 4-gluon transverse space
is given by
\beq
M^{(a)}(y)=\int\frac{dE}{2\pi}e^{-Ey}M^{(a)}_E
\eeq
and
\beq
M_E^{(a)}=\frac{1}{2}\Big(v_{13}+v_{24}-v_{23}-v_{14}\Big)
G_E^{(1243)}
\Big(v_{12}+v_{34}-v_{23}-v_{14}\Big),
\eeq
where $G_E^{1243}$ is the Green function of the BKP state with gluons
1,2,3 and 4 ordered as 1243.
The corresponding funtion $F$ obtained by the rules presented above
will be given by

\[
F_{4}^P=4\int_0^Y dy'\int_0^{y'}dy''
\int\prod_{j=1}^4\frac{d^2q'_j}{(2\pi)^2}\,\frac{d^2q_j}{(2\pi)^2}
2\pi\delta^2\Big(\sum_{j=1}^4q'_j-\sum_{j=1}q_j\Big)\]\[
<q_1,-q_1,-q_4,q_4|M^{(a)}(y'-y'')|q'_1,-q'_4,-q'_2,q'_4>\]\beq
P(Y-y',q_4)P(y'',q'_4)
\Big(P_{y,k}(Y-y',q_1)P(y'',q'_1)
+P(Y-y',q_1)P_{y,k}(y'',q'_1)\Big).
\label{f3p}
\eeq

The contribution from the diagram in Fig. \ref{fig1},5 will be given by
the same
expression with $M^{(a)}\to M^{(b)}$
\beq
F_{5}^P=F_{4}^P(M^{(a)}\to M^{(b)})
\label{fp5}
\eeq
where as a function of energy $E$
\beq
M_E^{(b)}=\Big(v_{13}+v_{24}-v_{23}-v_{14}\Big)
G_E^{(1234)}
\Big(v_{13}+v_{24}-v_{23}-v_{14}\Big).
\eeq

\section{Single interaction between the
pomerons attached to the projectiles and target}
\begin{figure}[h]
\leavevmode \centering{\epsfysize=0.3\textheight\epsfbox{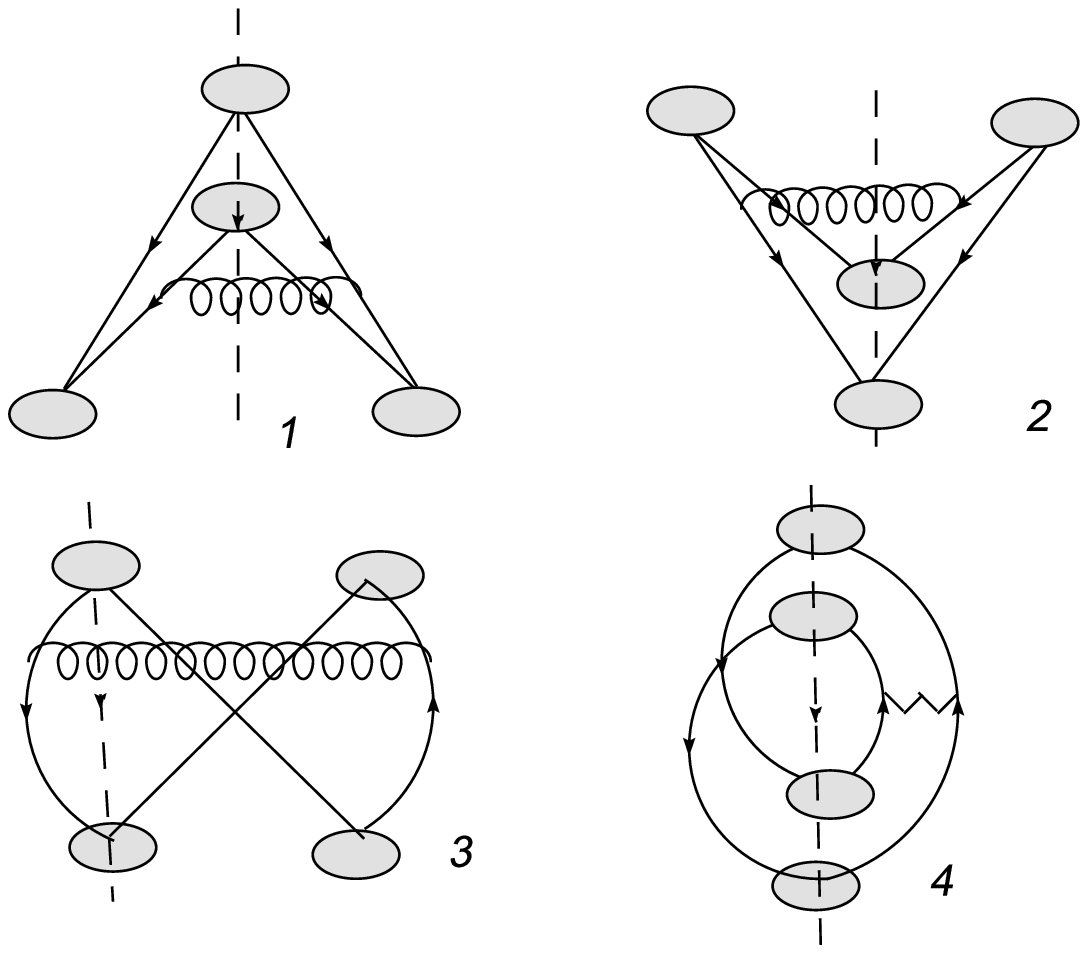}}
\caption{Different cuts in the amplitude shown in Fig. \ref{fig1},2 }
\label{fig4}
\end{figure}

Here we study the contribution to the inclusive cross-section which comes
from opening the interaction $H$ in the diagrams of Fig. \ref{fig1},
2 and 3. Obviously this implies substitution in Eq. (\ref{fina2})
\beq
<q,q'|H|q',q>\to <q,q'|v|q',q>(2\pi)^2\delta^2(q-q'-k)
\eeq
Again we shall have different contributions depending on the cuts in
the overall amplitude. They are shown in Fig. \ref{fig4},1-4.
Diagrams 1 and 2 describe the two D configurations, DT and DP,
diagrams 3 and 4
describe the S and DC configurations. Note that the DC contribution
Fig. \ref{fig4},4 does not contain the observed gluon in the
intermediate state and so gives no contribution. All the rest
contributions should be taken
with coefficient 2 due to interchanges of projectiles and targets.
So each can be presented in the form
\beq
F^V=4\kappa\int_0^ydy'\int\frac{d^2qd^2q'}{(2\pi)^2}
<q,q'|v|q',q>\delta^2(q-q'-k)
P(y-y',q)P(y-y',q')P(y',q)P(y',q')
\label{fgen}
\eeq
where the numerical coefficient $\kappa$ depends on a particular
configuration.
One trivially finds that
\beq
\kappa^{DT}=\kappa^{DP}=1,\ \ \kappa^{DC}=0,\ \ \kappa^S=-1
\eeq
In the sum the S contribution cancels half of the D contribution and
we obtain for the total contribution from a single interaction
\beq
F_V=4\int_0^ydy'\int\frac{d^2qd^2q'}{(2\pi)^2}
<q,q'|v|q',q>\delta^2(q-q'-k)
P(y-y',q)P(y-y',q')P(y',q)P(y',q')
\label{fv}
\eeq

\section{Contributions from two interactions between the
pomerons attached to the participants with
redistribution of colour}
\begin{figure}[h]
\leavevmode \centering{\epsfysize=0.5\textheight\epsfbox{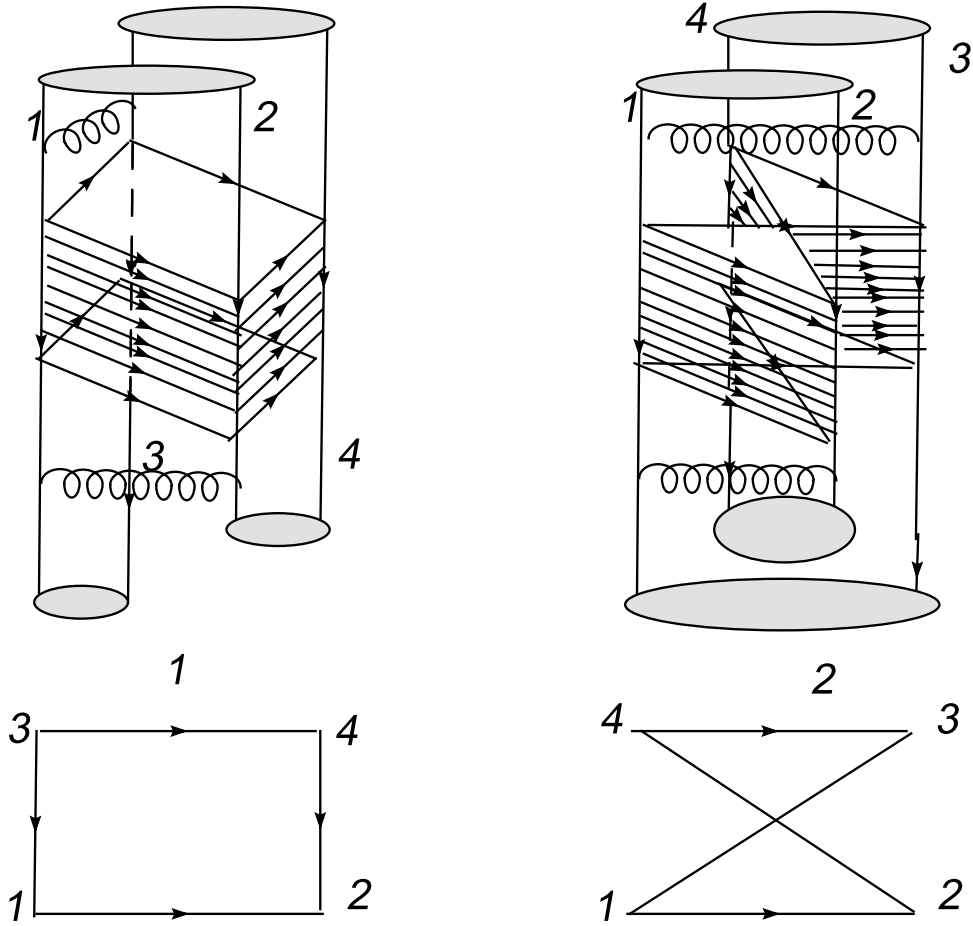}}
\caption{Contribution to the amplitude shown in Fig, \ref{fig1},4
with upper and lower interactions $v_{13}$ and $v_{12}$
respectively in two different forms with different order of
exchanged gluons. The schemes below show interactions between the
different gluons}
\label{fig5}
\end{figure}

\begin{figure}[h]
\leavevmode \centering{\epsfysize=0.3\textheight\epsfbox{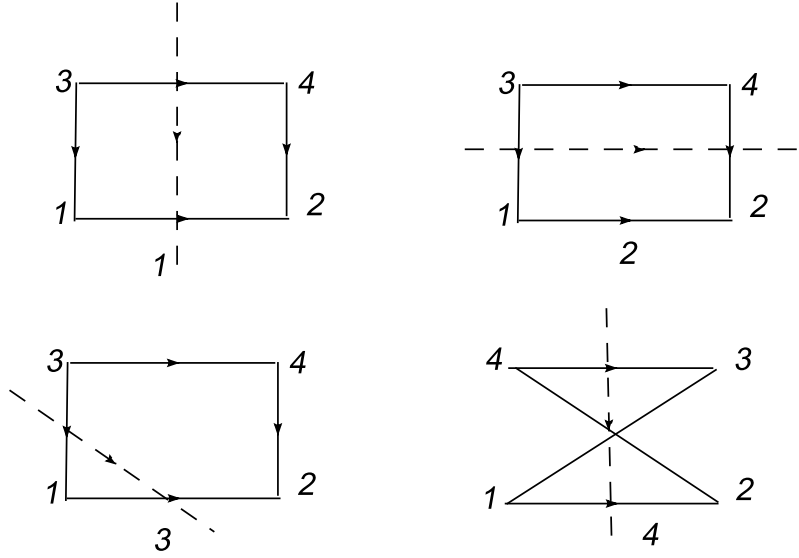}}
\caption{Cut interactions between different exchanged gluons for different
configurations}
\label{fig6}
\end{figure}

In this section we study contributions to the inclusive cross-section which
come from opening the interactions explicitly shown in Fig. \ref{fig1},4.
The amplitude itself is given by Eq. (\ref{a3ta}). The relevant cuts
are to pass through the interaction which contains the observed gluon.
They have also to pass through the diagram as a whole and thus have to
pass through some pomerons attached to the projectile and target and also
through the BKP state. The position of the latter cut will be totally
determined by the cut passing through the pomerons. To see this we have
to visualize the amplitude containing the BKP state in the form which
illustrates the intermediate $s$-channel (real) gluons in the whole
diagram. Unfortunately this can only be made in the 3-dimensional figure.
In Fig. \ref{fig5} we show one of the contributions to the amplitude
(\ref{a3ta}) in which the BKP state appears between interactions
$v_{13}$ and $v_{12}$ in two different forms with the gluons
in the Green function $G^{1243}$ placed on the surface of a cylinder
(1) and in the natural order 1234 (2).
In both cases the schemes below
indicate interactions betwen the gluons 1,2,3 and 4. This figure allows
to see how the cutting plane crosses the BKP state for different
configurations. In the diffractive configuration respective to
the target gluons 1 and 2 should lie on the opposite sides of the cut and so
should gluons 3 and 4. Gluons 1 and 3  as well as gluons 2 and 4
should lie on the same side of the cut plane. This means that in the
Fig. \ref{fig1},1 the cut plane should cross the BKP state as
indicated in Fig. \ref{fig6},1. For the diffractive contribution
respective to the projectile the cut plane should cross the BKP state
as indicated in Fig. \ref{fig6},2. For single cut contributions
one of the gluons should lie on one side of the cut plane and the three
others on the opposite side, as shown in Fig. \ref{fig6},3.
Finally the double cut contribution requires that, say, gluons 1 and 4
lie on one side of the cut and gluons 2 and 3 on the opposite side.
This requires the cut plane pass through the BKP state as shown in
Fig. \ref{fig6},4. The important point is that for each configuration the
cutting plane crosses the BKP state in the unique well-defined manner.
Since cut and uncut interactions give the same contribution, this also
means that in all cases the BKP state will appear as a whole between
the interactions which generate the observed gluon. As a result, to
study contributions from the explicitly shown interactions it is sufficient
to consider the situation when there are no interactions in the BKP state,
that is when $G^{1243}(y)\to 1$, and  introduce this  Green
function between the interactions afterwards.

With $G^{1243}(y)\to 1$ the amplitude simplifies to
\[
D_{3a}=\frac{1}{2}\int_0^Y dy'\int_0^{y'}dy''
\int d\tau^3_\perp
P_{12}(Y-y')P_{34}(Y-y')\]\beq
\Big(v_{13}+v_{24}-v_{23}-v_{14}\Big)
\times\Big(v_{12}+v_{34}-v_{23}-v_{14}\Big)
P_{13}(y'')P_{24}(y'')
\label{a3ta1}
\eeq
where we indicated numbers of the gluons but suppressed
transverse momenta
in both the pomerons and interactions between the gluons,
assuming that the interactions act on the initial pomerons $P_{12}$
and $P_{34}$
on the left and
on the final pomerons $P_{13}$ and $P_{24}$ on the right.
Integration in the transverse space goes over three independent
tranverse momenta.

The inclusive cross-sections are obtained by opening either the
upper interactions, which implies fixing $Y-y'=y$ and the momenta
transferred to $k$ in one of the 4 interactions on the left in
Eq. (\ref{a3ta1}), or the lower interactions, which implies fixing
$y''=y$ and the momentum transferred in one of the interactions on the
right in Eq. (\ref{a3ta1}).

In both cases contributions will
come from DT, DP, S and DC configurations.
We shall write our results for the high-energy part $F$  in the form
\beq
F_{high}^{2V}= \int_0^ydy'' d\tau^2_\perp
P_{12}(Y-y)P_{34}(Y-y)f_{high}P_{13}(y'')P_{24}(y'')
\label{ffh}
\eeq
for emission from the interaction higher in rapidity and
\beq
F_{low}^{2V}= \int_y^{Y}dy' d\tau^2_\perp
P_{12}(Y-y')P_{34}(Y-y')f_{low}P_{13}(y)P_{24}(y)
\label{ffl}
\eeq
for emission from the  interaction lower in rapidity.
Here $f_{high}$ and $f_{low}$ are certain operators which are different for
different configurations.
In all cases contributions have to be multplied by 4 for D and S
configurations and by 2 for the DC configuration. These factors
will be included in the overall factors in (\ref{ffh}) and
(\ref{ffl}).

\begin{figure}[h]
\leavevmode \centering{\epsfysize=0.3\textheight\epsfbox{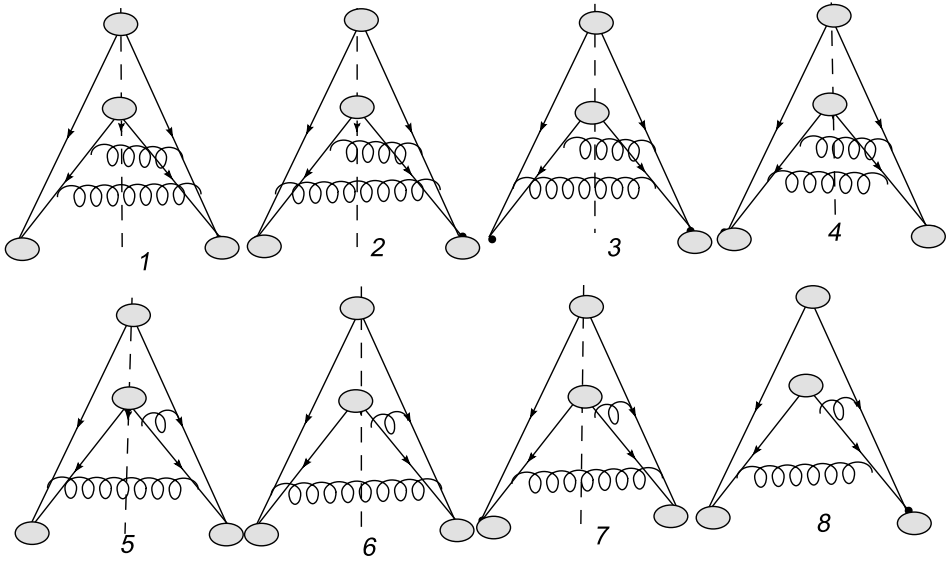}}
\caption{Diffractive contribution with respect to the targets}
\label{fig7}
\end{figure}

\begin{figure}[h]
\leavevmode \centering{\epsfysize=0.3\textheight\epsfbox{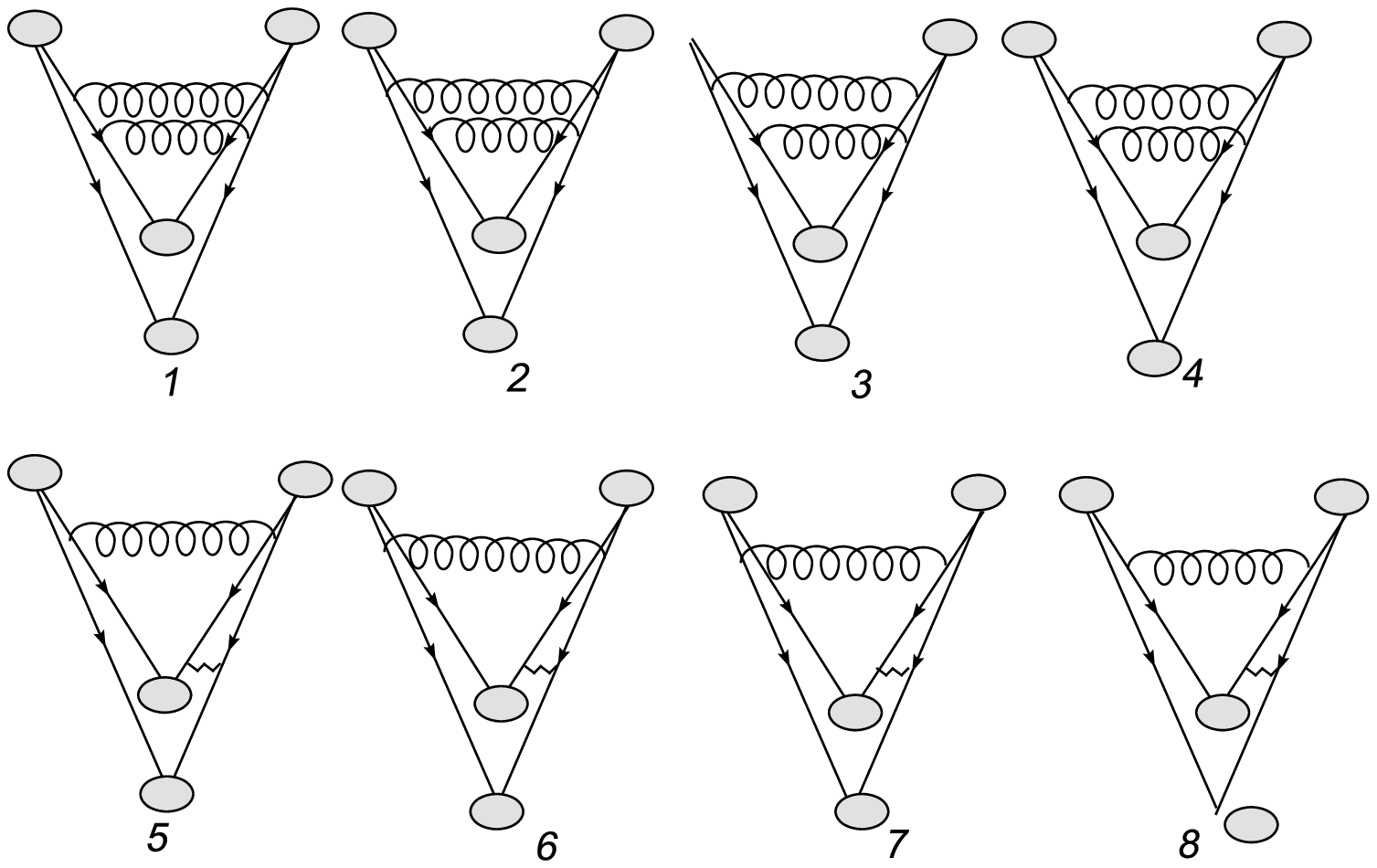}}
\caption{Diffractive contribution with respect to the projectile}
\label{fig8}
\end{figure}

We start with emission from the upper interaction.
The contribution diffractive with respect to the targets
with two gluons in the intermediate state (DT2), Fig. \ref{fig7},1-4, is
\beq
f_{DT2}=
v_{23}v_{23}-v_{23}v_{12}+v_{23}v_{14}-v_{23}v_{34}.
\label{fdt2}
\eeq
The contribution diffractive with respect to the projectiles
 with two gluons in the intermediate state (DP2), Fig.\ref{fig8},1-4, is
\beq
f_{DP2}=
v_{23}v_{23}-v_{13}v_{23}+v_{14}v_{23}-v_{24}v_{23}.
\label{fdp2}
\eeq
The D contribution respective to the projectile
with one gluon in the intermediate state (DP1), Fig. \ref{fig8},5-8, is
\beq
f_{DP1}=
-v_{23}v_{34}-v_{14}v_{34}+v_{13}v_{34}+v_{24}v_{34}.
\label{fd1}
\eeq

\begin{figure}[h]
\leavevmode \centering{\epsfysize=0.15\textheight\epsfbox{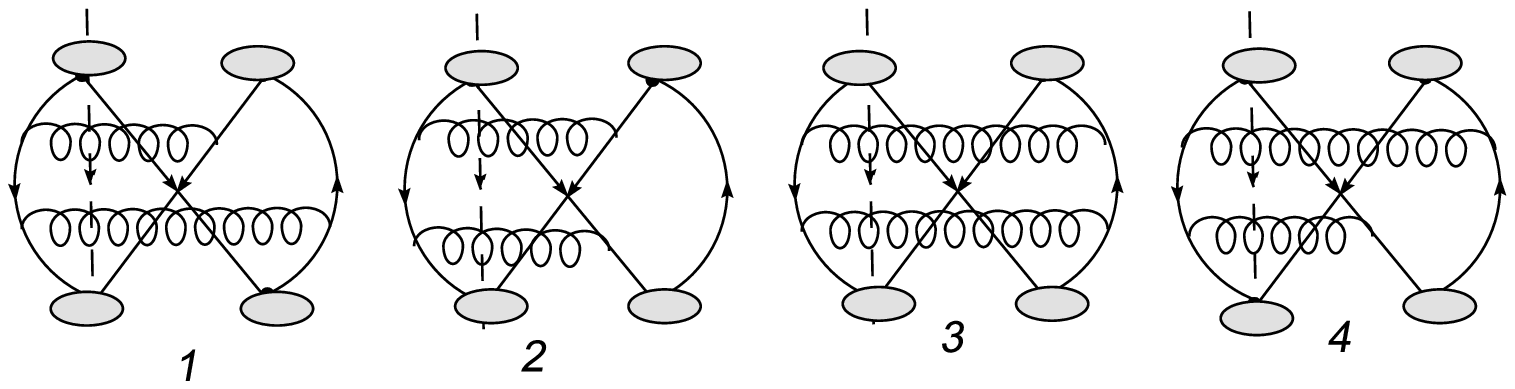}}
\caption{Single cut contribution with two gluons in the intermediate state }
\label{fig9}
\end{figure}

\begin{figure}[h]
\leavevmode \centering{\epsfysize=0.3\textheight\epsfbox{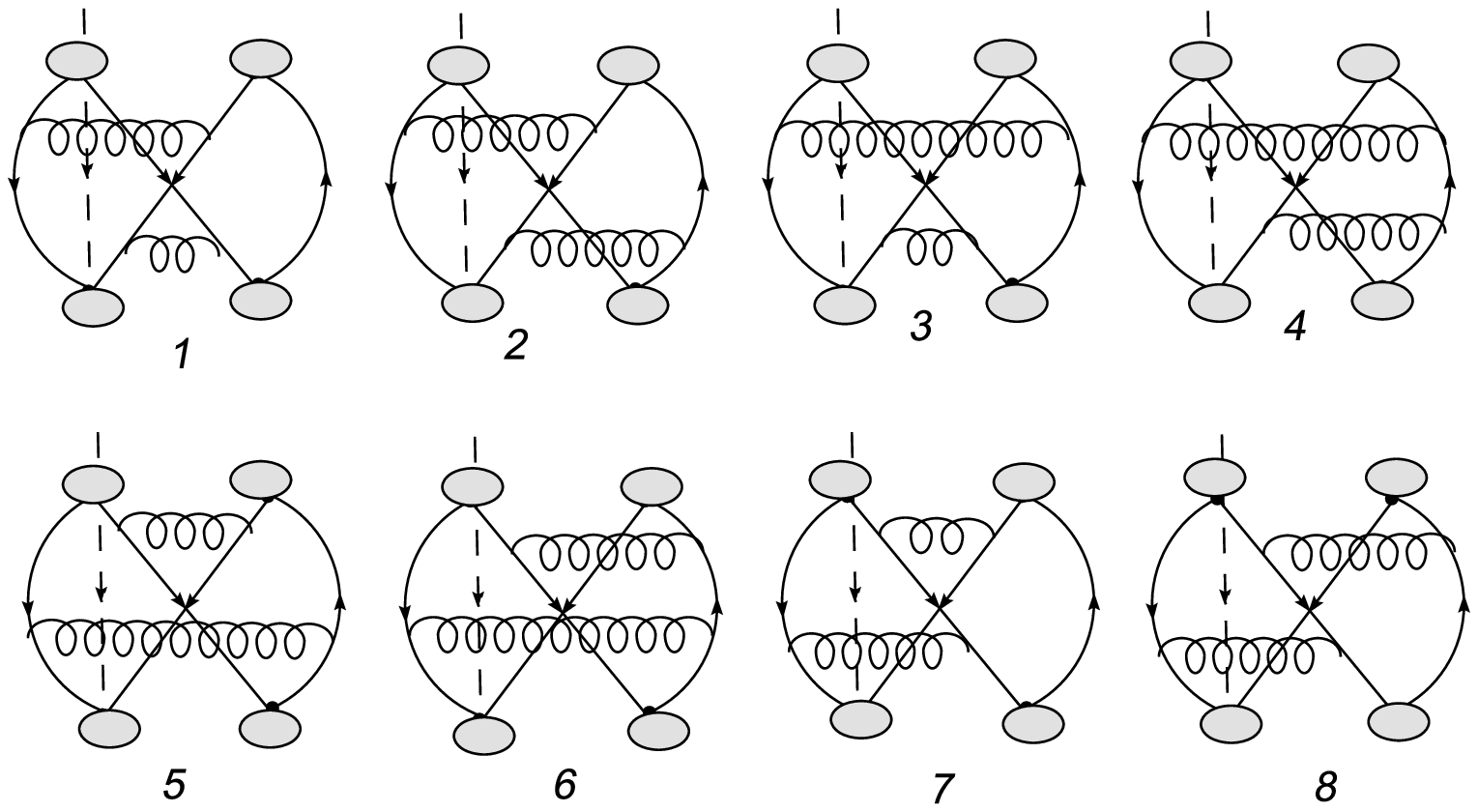}}
\caption{Single cut contribution with one gluon in the intermediate state }
\label{fig10}
\end{figure}

\begin{figure}[h]
\leavevmode \centering{\epsfysize=0.3\textheight\epsfbox{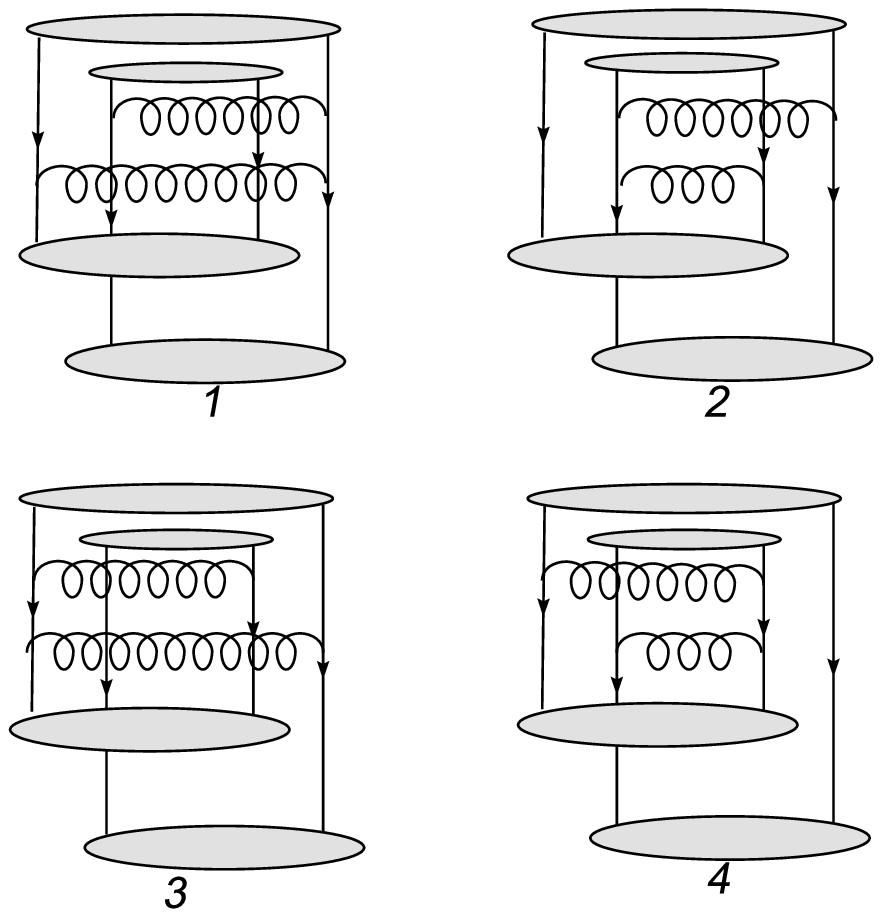}}
\caption{Double  cut contributions with two gluons in the intermediate state }
\label{fig11}
\end{figure}

\begin{figure}[h]
\leavevmode \centering{\epsfysize=0.3\textheight\epsfbox{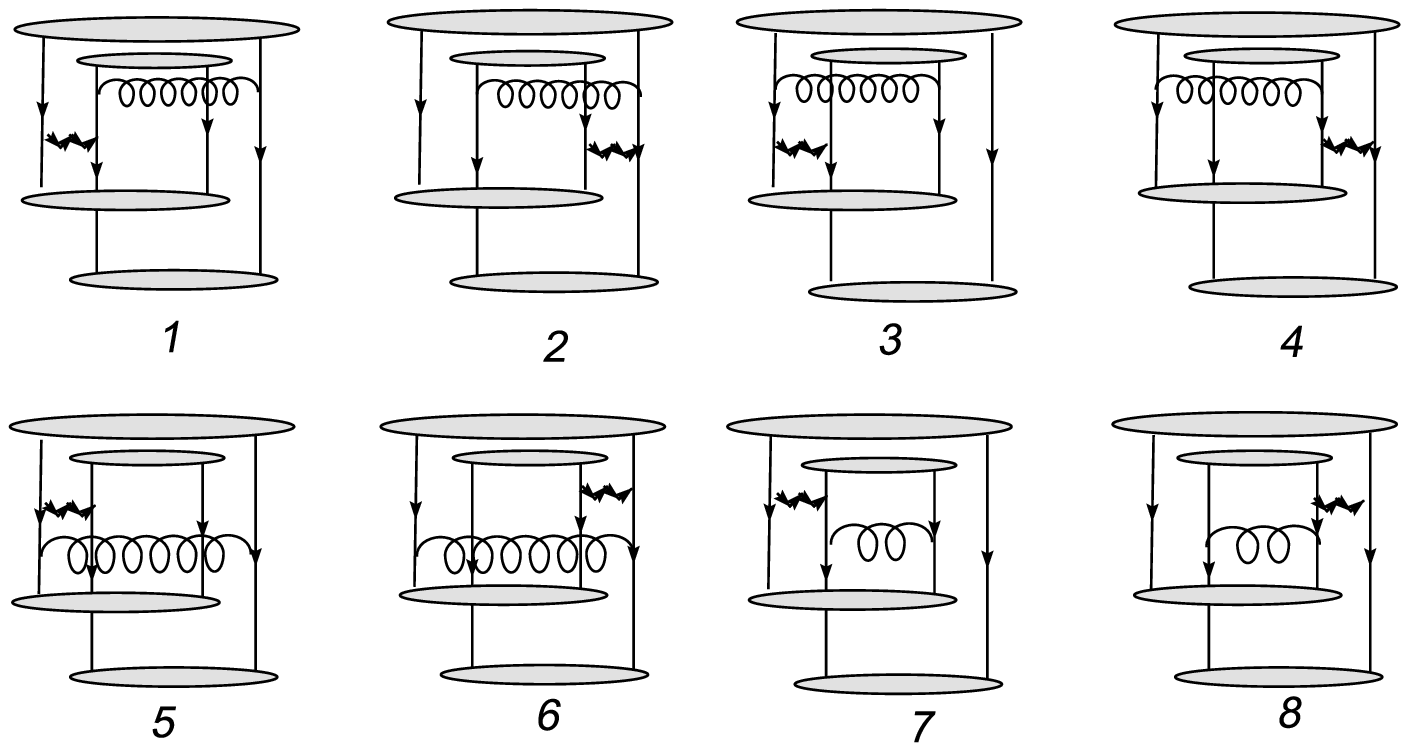}}
\caption{Double  cut contribution with a single gluon in the intermediate state }
\label{fig12}
\end{figure}

The S contribution with two gluons in the intermediate state (S2),
Fig. \ref{fig9}, is
\beq
f_{S2}=
v_{13}v_{14}-v_{13}v_{12}-v_{23}v_{23}+v_{23}v_{12}.
\label{fs2}
\eeq
The S contribution with one gluon in the intermediate state (S1).
Fig. \ref{fig10}, is
\beq
f_{S1}=
v_{13}v_{23}-v_{13}v_{34}-v_{14}v_{23}+v_{14}v_{34}.
\label{fs1}
\eeq
The DC contributions with both two gluons and one gluon in the
intermediate state (DC=DC1+DC2), Figs. \ref{fig11} and \ref{fig12} is
\beq
f_{DC}=
v_{24}v_{12}+v_{24}v_{34}-v_{24}v_{14}-v_{24}v_{23}.
\label{DC}
\eeq
Summing these contributions we find
\[
f_{high}=(v_{23}+v_{14}-v_{13}-v_{24})(v_{23}-v_{34}).
\]
The observed gluon is emitted from the interaction on the left.
Obviously this contribution does not contain a pole at $p'=0$ and so
is infrared safe.

Now we consider contributions with the observed gluon of a lower rapidity
The D contributions with two gluons in the intermediate state are
\beq
f_{DT2}=
v_{23}v_{23}-v_{23}v_{12}+v_{23}v_{14}-v_{23}v_{34},
\label{fd2l}
\eeq
\beq
f_{DP2}=
v_{23}v_{23}-v_{13}v_{23}+v_{14}v_{23}-v_{24}v_{23},
\label{fdp2l}
\eeq
The D contribution with one gluon in the intermediate state
(from diffraction relative to the targets) is
\beq
f_{DT1}=
-v_{24}v_{23}+v_{24}v_{12}-v_{24}v_{14}+v_{24}v_{34}.
\label{fd1l}
\eeq
The S contribution with two gluons in the intermediate state is
\beq
f_{S2}=
v_{13}v_{14}-v_{13}v_{12}-v_{23}v_{23}+v_{23}v_{12}.
\label{fs2l}
\eeq
The S contribution with one gluon in the intermediate state is
\beq
f_{S1}=
-v_{23}v_{14}+v_{24}v_{14}+v_{23}v_{12}-v_{24}v_{12}.
\label{fs1l}
\eeq
The DC contribution with both two gluons and one gluon in the
intermediate state is
\beq
f_{DC}=
v_{24}v_{12}+v_{24}v_{34}-v_{14}v_{12}-v_{23}v_{34}.
\label{fdcl}
\eeq
Summing these contributions we find
\[
f_{low}=(v_{23}+v_{14}-v_{13}-v_{24})(v_{23}-v_{34}).
\]
Remarkably it has the same form as for $y>y'$. However in this case the
observed gluon is emitted from the interaction on the right. Again the
poles at $p'=0$ in the left factor cancel. So this contribution
is also infrared safe.

Note that we have an obvious symmetry from the identity
of the two projectiles and two targets: simultaneous interchange
$1\lra 2$ and $3\lra 4$ does not change the result.
Apart from this symmetry
our diagrams possess symmetry $(1,2)\lra(4,3)$. Usung these
two symmetries we can transform our expression to  find
the result for $f$,
valid both for $y>y'$ and $y<y'$
\beq
f=\frac{1}{2}(v_{23}+v_{14}-v_{13}-v_{24})
(v_{23}+v_{14}-v_{12}-v_{34}).
\label{ffin}
\eeq
It coincides with the corresponding form of $f$ for the total
cross-section, which was to be expected, since with a single
gluon inside the interaction integration over $y$ and $k$ should
give twice the coresponding part of the total ceos-section.

Inclusion of the nontrivial Green function for the BKP state will
give our final expression for the
inclusive cross-section from the two explicit interactions in the diagram
Fig. \ref{fig1},4
\beq
f=\frac{1}{2}(v_{23}+v_{14}-v_{13}-v_{24})G^{1243}(y'-y'')
(v_{23}+v_{14}-v_{12}-v_{34}).
\label{ffin1}
\eeq
Here the observed gluon may be located either in the left
interaction or the right one and correspondingy $y'=y$ or $y''=y$.

\section{Contributions from two interactions between the
pomerons attached to the participants with
direct colour transmission (DCT)}
In this section we study contributions to the inclusive
cross-section which
come from opening the interactions explicitly shown in Fig.
\ref{fig1},5.
In this case we  enumerate the gluons
connecting  the first projectile with the first target  1 and 2,
and those connecting the second projectile with the second target as
3 and 4,
Note that diagrams with interactions involving gluons 1 and 2 are to be
considered as different, which is clear when the original diagrams
with quark propagators are studied. The same is valid for gluons
3 and 4. In this case  the identity
of the two projectiles and two targets generates symmetries in
independent interchanges
$1\lra 2$ or $3\lra 4$.

\begin{figure}[h]
\leavevmode \centering{\epsfysize=0.5\textheight\epsfbox{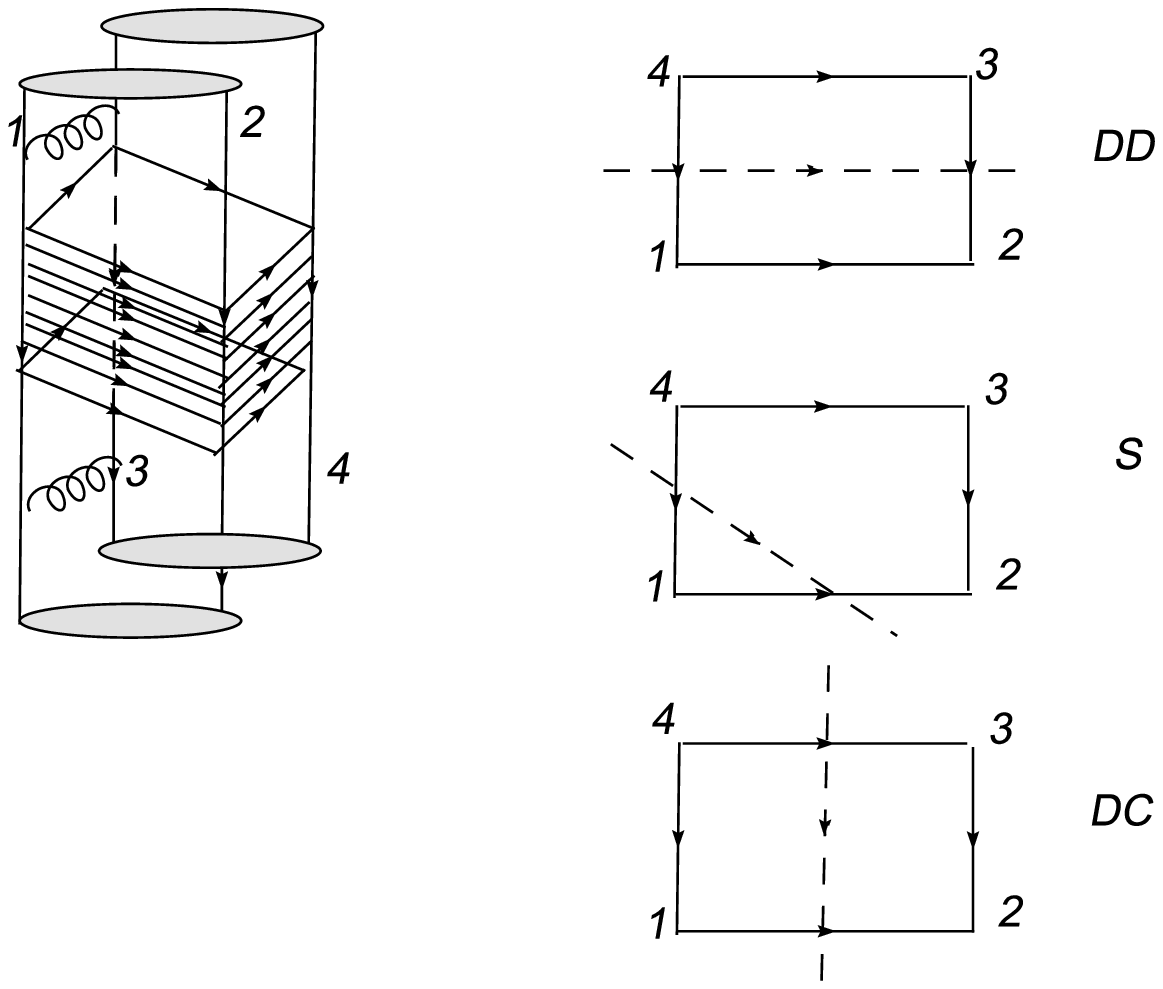}}
\caption{Contribution to the amplitude shown in Fig, \ref{fig1},5
with both upper and lower interactions $v_{13}$.
 The schemes on the right show interactions between the
different gluons for different cut configurations}
\label{fig13}
\end{figure}

The amplitude itself is given by Eq. (\ref{a3ta}) with the substitution
$M^{(a)}\to M^{(b)}$. As before the relevant cuts
are to pass  through the interaction which contains the observed
gluon.
and also through the diagram as a whole and thus
through the BKP state. The position of the latter cut will again
be totally
determined by the cut passing through the pomerons, which can be
seen from Fig. \ref{fig13}, in which we
 show one of the contributions to the amplitude
 with the BKP state appearing between two interactions
$v_{13}$. In the schemes below indicating
interactions betwen the gluons 1,2,3 and 4 we show how the BKP state
is cut.
This figure allows
to see how the cutting plane crosses the BKP state for the three
possible configurations: diffractive both with respect to the targets
and projectiles (DD), single cut (S) an double cut (DC).
Again for each configuration the
cutting plane crosses the BKP state in the unique well-defined manner.
Since, as mentioned, cut and uncut interactions give the same
contribution, in all cases the BKP state will appear as a whole between
the interactions which generate the observed gluon. As a result, to
study contributions from the explicitly shown interactions, as before,
it is sufficient
to consider the situation when there are no interactions in the BKP state,
that is when $G^{1234}(y)\to 1$ and afterwards introduce this  Green
function between the interactions.

With $G^{1234}(y)\to 1$ the amplitude simplifies to
\[
D_{3b}=\frac{1}{2}\int_0^Y dy'\int_0^{y'}dy''
\int d\tau^3_\perp
P_{12}(Y-y')P_{34}(Y-y')
\Big(v_{13}+v_{24}-v_{23}-v_{14}\Big)\]\beq
\times\Big(v_{13}+v_{24}-v_{23}-v_{14}\Big)
P_{13}(y'')P_{24}(y'')
\label{a3tb1}
\eeq
in the same notation as in the previous section.

\begin{figure}[h]
\leavevmode \centering{\epsfysize=0.3\textheight\epsfbox{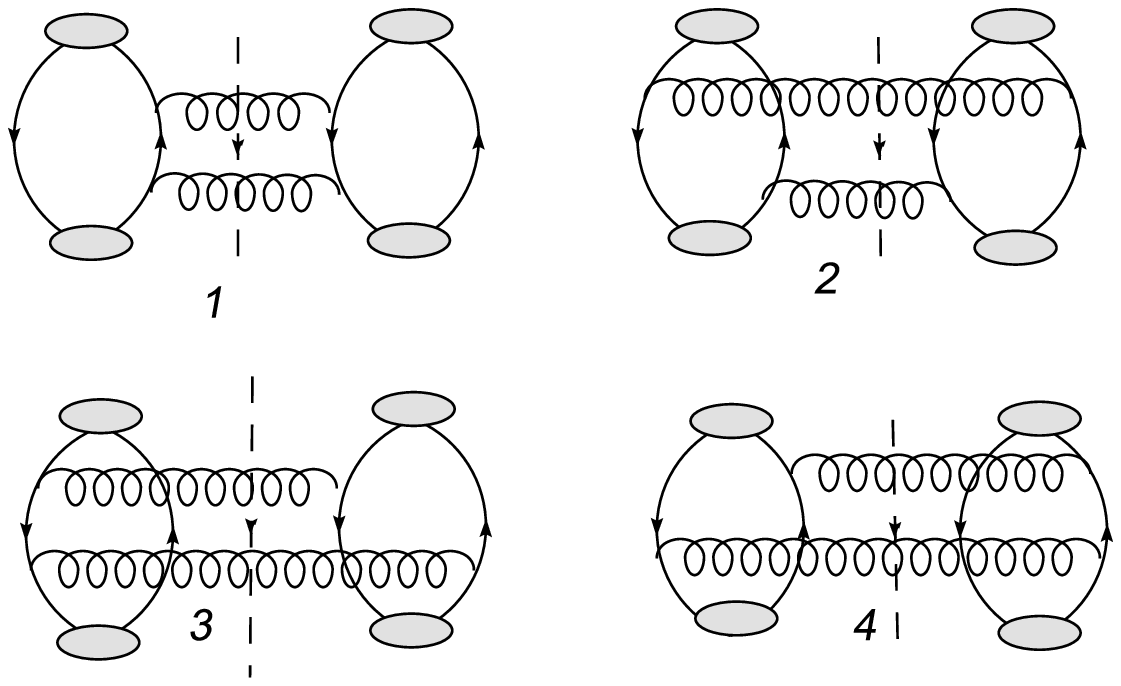}}
\caption{Typical diagrams for the double diffractive contribution.
Only one fourth of the total number 16 of diagrams are shown}
\label{fig14}
\end{figure}

\begin{figure}[h]
\leavevmode \centering{\epsfysize=0.3\textheight\epsfbox{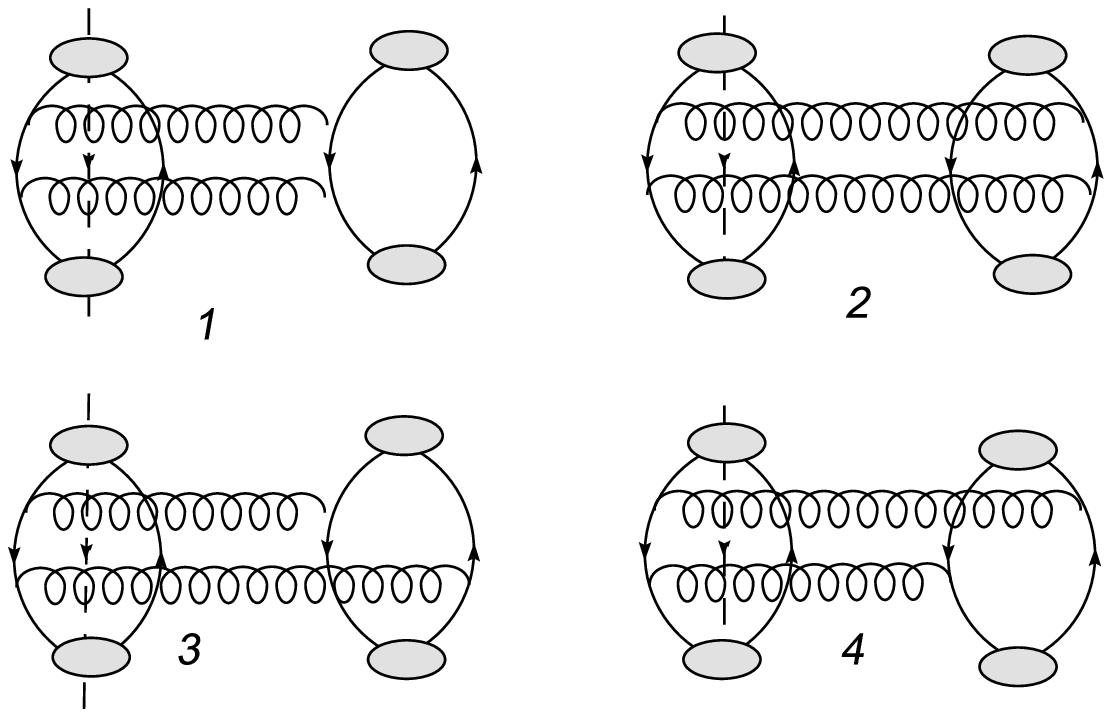}}
\caption{Single cut contribution for DCT with two gluons in the
intermediate state. }
\label{fig15}
\end{figure}

\begin{figure}[h]
\leavevmode \centering{\epsfysize=0.3\textheight\epsfbox{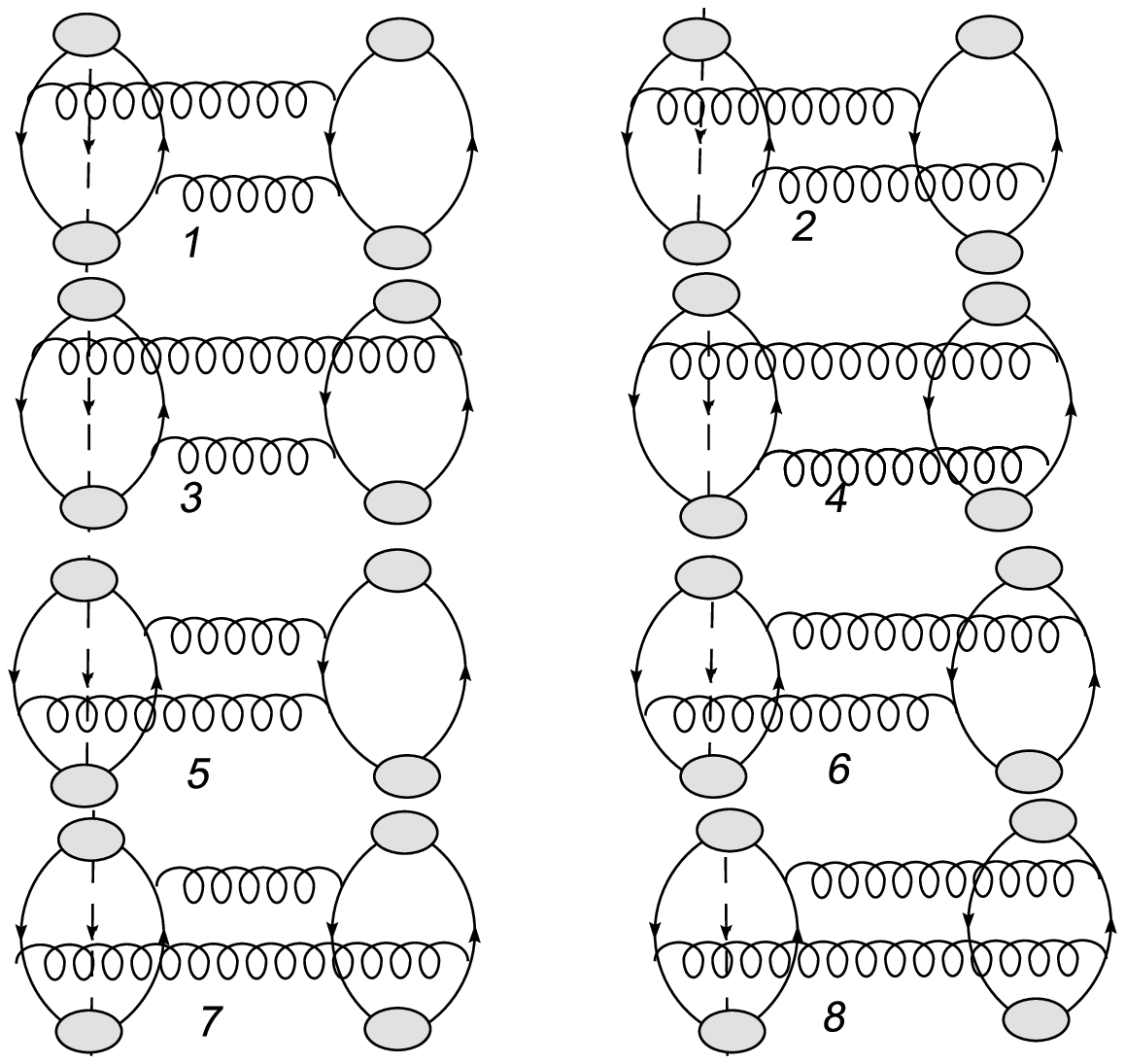}}
\caption{Single cut contribution for DCT with one gluon in the
intermediate state.}
\label{fig16}
\end{figure}

The inclusive cross-sections are obtained by opening either the
upper interactions, which implies fixing $Y-y'=y$ and the momenta
transferred to $k$ in one of the 4 interactions on the left in
Eq. (\ref{a3tb1}), or the lower interactions, which implies fixing
$y''=y$ and the momentum transferred in one of the interactions on the
right in Eq. (\ref{a3tb1}).
In both cases contributions will
come from DD,  S and DC configurations.
We shall write our results for the high-energy part $F$  in the
same forms as before
\beq
\tilde{F}_{high}^{2V}=\int_0^ydy'' d\tau^2_\perp
P_{12}(Y-y)P_{34}(Y-y)\tf_{high}P_{13}(y'')P_{24}(y'')
\label{tffh}
\eeq
for emission from the upper interaction
and
\beq
\tilde{F}_{low}^{2V}= \int_y^{Y}dy' d\tau^2_\perp
P_{12}(Y-y')P_{34}(Y-y')\tf_{low}lP_{13}(y)P_{24}(y)
\label{tffl}
\eeq
for emission from the lower interaction.
As before contributions for all $F$'s are to be taken with coefficient
4, except for DC contributions which are to be taken with coefficient 2.
For the observed gluon
of a higher rapidity we find from our formulas
the double diffractive (DD) contribution, Fig. \ref{fig14} as
\beq
\tf_{DD}=2v_{23}(v_{23}+v_{14}-v_{13}-v_{24}).
\label{bfdd}
\eeq
The S contribution with two gluons in the intermediate state
Fig. \ref{fig15}
\beq
\tf_{S2}=
2v_{13}(v_{14}-v_{13}).
\label{bfs2}
\eeq
The S contribution with one gluon in the intermediate state
Fig. \ref{fig16}
\beq
\tf_{S1}=
2v_{13}(v_{23}-v_{24}).
\label{bfs1}
\eeq

\begin{figure}[h]
\leavevmode \centering{\epsfysize=0.2\textheight\epsfbox{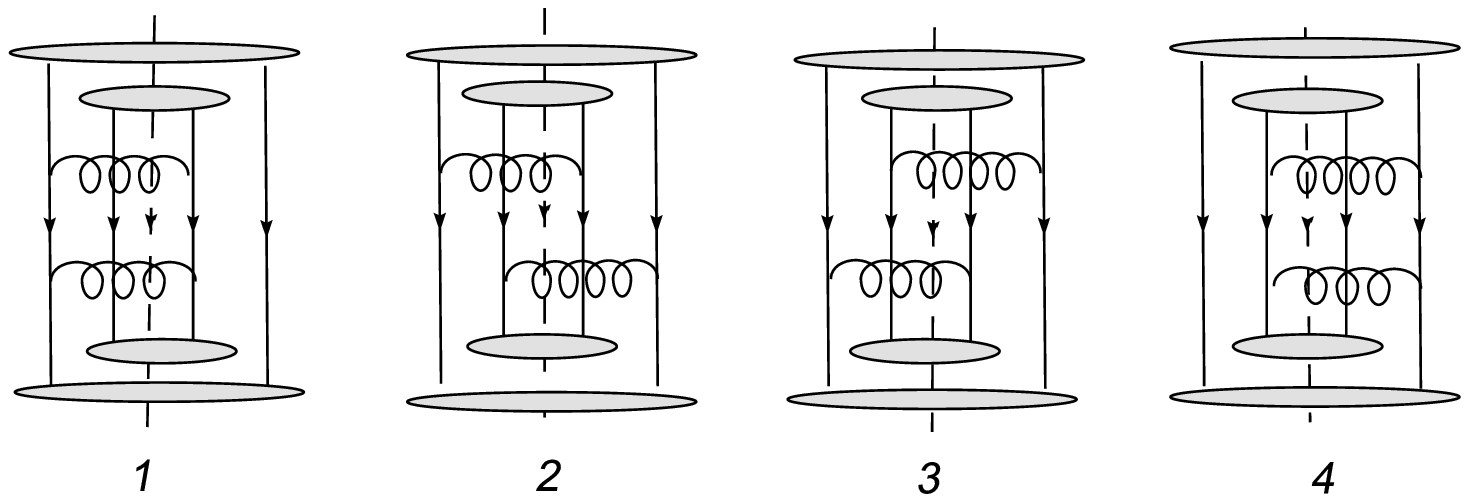}}
\caption{Double cut contribution for DCT with two gluons in the
intermediate state. }
\label{fig17}
\end{figure}

\begin{figure}[h]
\leavevmode \centering{\epsfysize=0.3\textheight\epsfbox{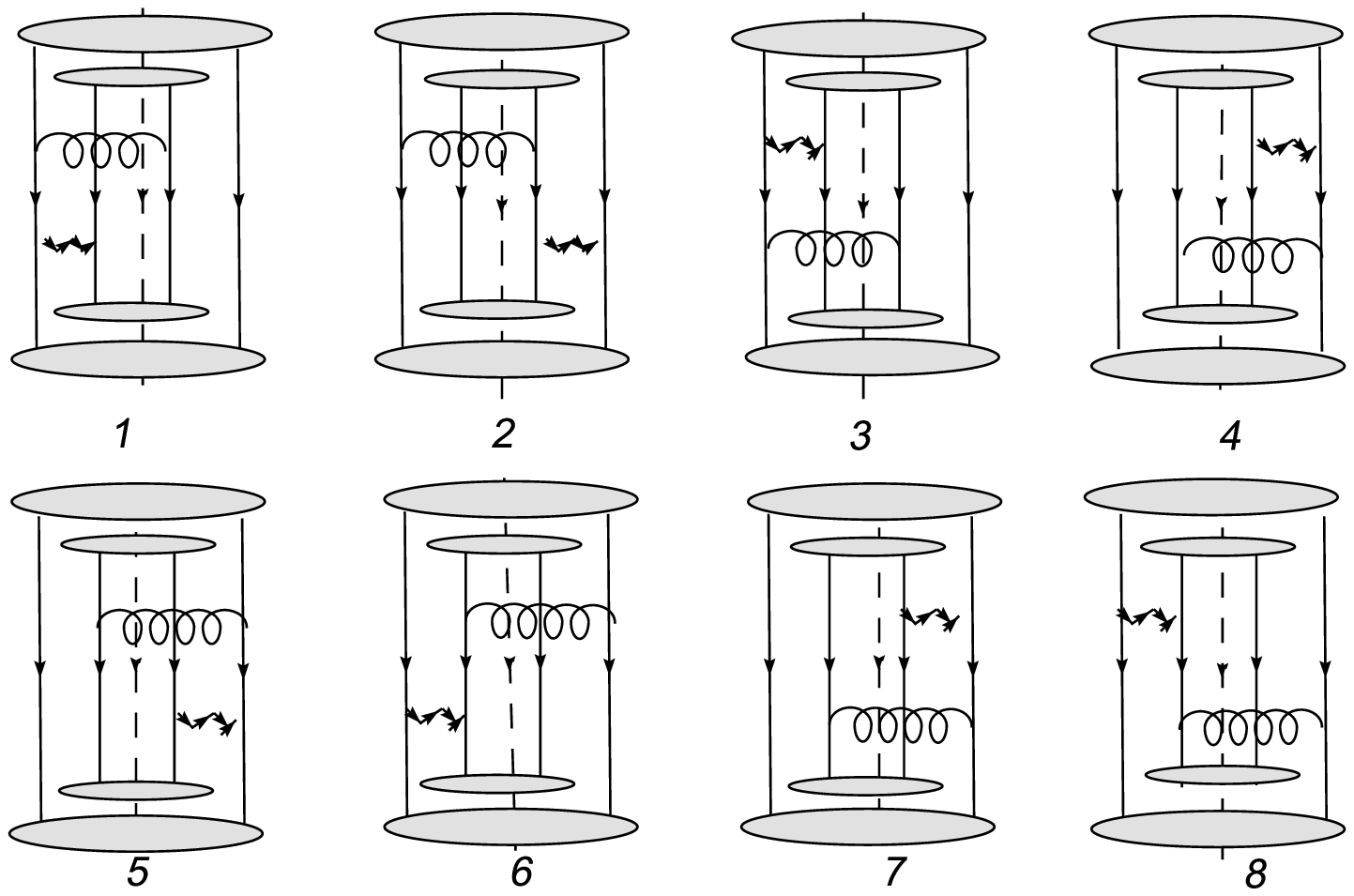}}
\caption{Double cut contribution for DCT with one gluon in the
intermediate state.}
\label{fig18}
\end{figure}

The DC contribution with both two gluons and one gluon in the
intermediate state Fig. \ref{fig17} and \ref{fig18}
\beq
\tf_{DC}=
2v_{14}(v_{14}+v_{23}-v_{13}-v_{24})=
2v_{13}(v_{13}+v_{24}-v_{14}-v_{23}).
\label{bfdc}
\eeq
We observe that $\tf_{S2}+\tf_{S1}+\tf_{DC}=0$. So the total
contribution is given by $\tf_{DD}$. It can be rewritten
as
\beq
\tf=\frac{1}{2}(v_{14}+v_{23}-v_{13}-v_{24})
(v_{14}+v_{23}-v_{13}-v_{24})
\label{bffin}
\eeq
and taking into account the intermediate BKP state
\beq
\tf=\frac{1}{2}(v_{14}+v_{23}-v_{13}-v_{24})G^{1234}(y-y')
(v_{14}+v_{23}-v_{13}-v_{24}),
\label{bffin1}
\eeq
which again coincides with the corresponding expression for the
total cross-section.

For the case when the observed gluon has a lower rapidity we have
to integrate $\tf$ according to Eq. (\ref{tffl}).
The expressions for $\tf_{DD}$, $\tf_{S2}$ and $\tf_{DC2}$ obviously
do not change. For $\tf_{S1}$ we find
\[\tf^{S1}=2v_{23}(v_{13}-v_{14})\]
However interchange $1\lra 2$ makes it equal to (\ref{bfs1}).
The contribution to $\tf_{DC}$ with a single gluon in the intermediate
state is
\[-2v_{13}v_{14}-2v_{24}v_{14}=
-2v_{14}(v_{13}+v_{24})\]
and is equal to the last two terms in (\ref{bfdc}).
So the total expression for $\tf$ is the same as for the case $y>y'$,
Eq. (\ref{bffin}).

\section{Contributions from  the BKP state}
\begin{figure}[h]
\leavevmode \centering{\epsfysize=0.2\textheight\epsfbox{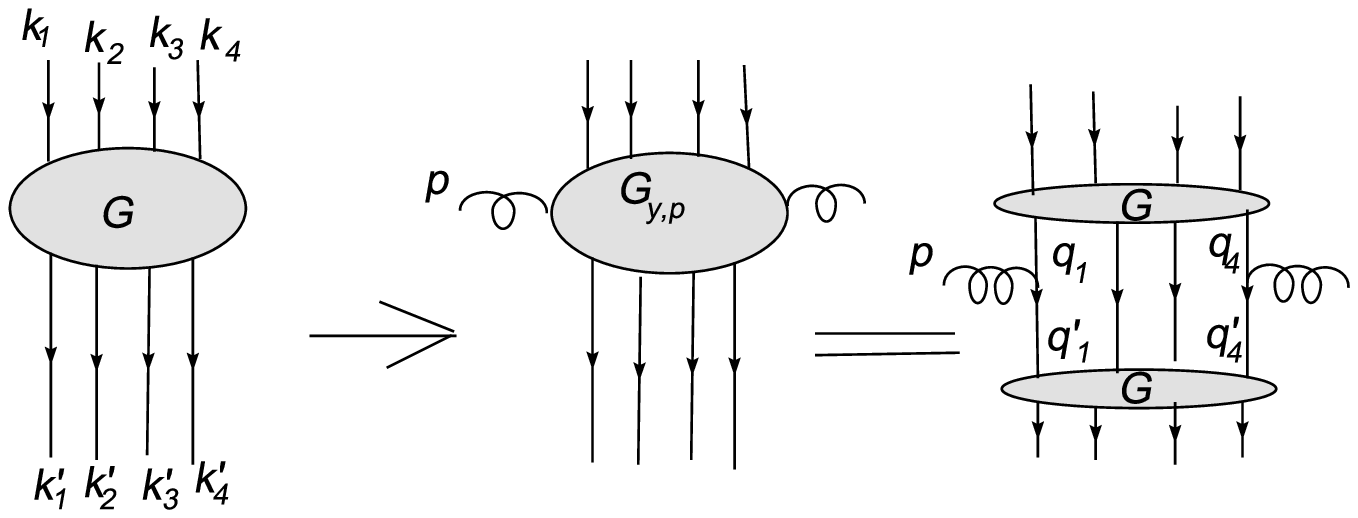}}
\caption{'Opening' the BKP state }
\label{fig19}
\end{figure}

By its structure the BKP state is very similar to the standard
BFKL pomeron. To fix one of the intermediate real gluons one has to 'open'
the BKP state similarly to opening the BFKL pomeron. This amounts
to presenting the BKP Green function as a convolution of two such
Green functions with the BFKl interaction containing the fixed real gluon
between them, that is  substitution $G\to G_{y,k}$ where
explicitly
\[
<k_1,k_2,k_3,k_4|G_{y,k}(y',y'')|k'_1k'_2k'_3k'_4> =
\int \prod_{i=1}^4\frac{d^2q_i}{(2\pi)^2}\frac{d^2q'_i}{(2\pi)^2}
\]\[
<k_1,k_2,k_3,k_4|G(y'-y)|q_1,q_2,q_3,q_4>
<q'_1,q'_2,q'_3,q'_4|G(y-y'')|k'_1k'_2k'_3k'_4>\]\[
(2\pi)^2\delta^2(\sum q_i)(2\pi)^2\delta^2(\sum q'_i)
\bar{v}(q_1,q_4|q'_1 q'_4)\]\beq
(2\pi)^2\delta^2(q_1+q_4-q'_1-q'_4)
(2\pi)^2\delta^2(q_1-q'_1-k)(2\pi)^2\delta^2(q_3-q'_4).
\label{gyp}
\eeq
This is illustrated in Fig. \ref{fig19}.

The choice of the interaction
with the observed gluon ($v_{14}$ in (\ref{gyp})) is of course arbitrary
due to the symmetry of $G^{1234}$. However there is one essential difference
between opening the BFKL pomeron and the BKP state. As we have observed
before, the cutting plane always passes through 2 or 4 interactions
connecting different gluons inside
the BKP state. Correspondingly the contribution from the BKP state has to be
multiplied by 2 or 4 depending on the number of crossed
interactions between different gluons. In fact we have seen that this
number is just 2 in all cases except the double cut configuration with
colour redistribution, when it is equal 4.

With these preliminary considerations it is
trivial to write the contributions for gluon production from the
BKP state illustrated in Fig. 20, 1 for colour redistribution and 2
with no colour redistribution.
\begin{figure}[h]
\leavevmode \centering{\epsfysize=0.25\textheight\epsfbox{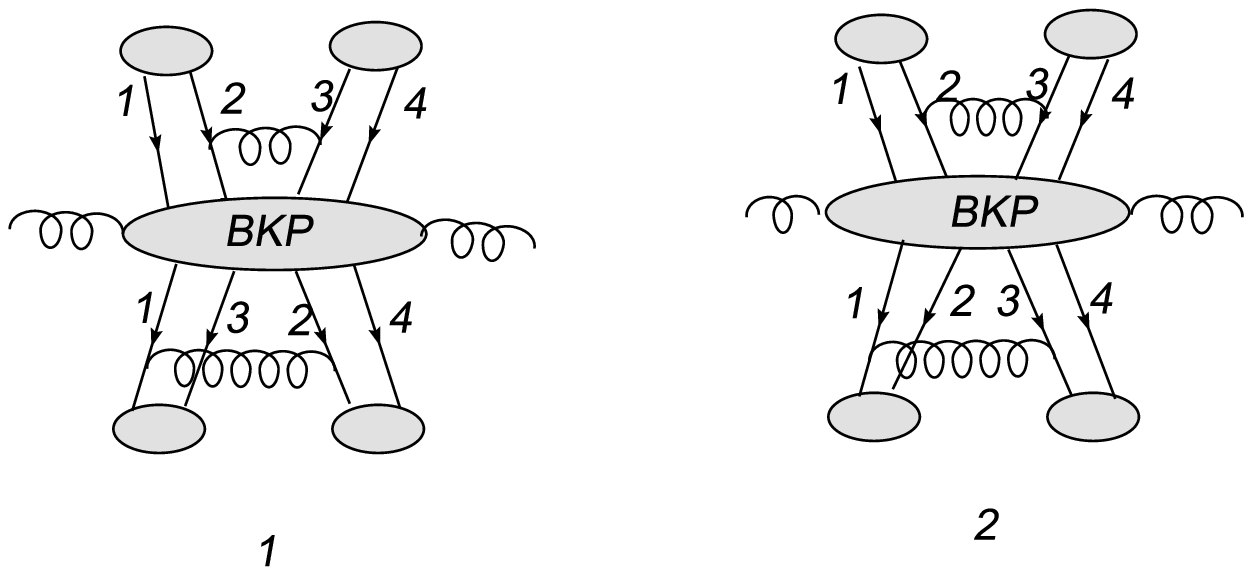}}
\caption{Gluon production from the
BKP state }
\label{fig20}
\end{figure}

We present the results in the form similar to the ones before.
For the case with colour redistribution
\beq
F^{BKP}= \int_0^Y dy'\int_0^{y'}dy'' d\tau^2_\perp
P_{12}(y')P_{34}(y')f(y',y'',y)P_{13}(y'')P_{24}(y'').
\label{fbkp}
\eeq
and for case without colour redistribution $F^{BKP}\to \tilde{F}^{BKP}$
and $f\to\tf$.
As before all contributions are to be taken with coefficient 4, except
the DC contribution which is to be taken with coefficient 2. These
coefficients are taken into account in the overall coefficient
in (\ref{fbkp}).

For the case with colour redistribution,
Fig. \ref{fig20},1 we find
in the D configuration
\beq
f_D=(v_{23}+v_{14}-v_{13}-v_{24})G^{1243}_{y,k}(y',y'')
(v_{23}+v_{14}-v_{12}-v_{34}).
\label{fbkpd}
\eeq
In the S configuration
\beq
f_S=-f_D
\eeq
and in the DC configuration
\beq
f_{DC}=f_D.
\eeq
Note that the twice lower factor for the DC contribution is
compensated by the twice large number of the interactions
connecting different gluons. Thus the final result for the
total contribution is just
$f(y,k)=f_D$

For the case without colour redistribution
Fig. \ref{fig20},2 we find
in the D configuration
\beq
\tf_D=(v_{23}+v_{14}-v_{13}-v_{24})G^{1234}_{y,k}(y',y'')
(v_{23}+v_{14}-v_{13}-v_{24}).
\label{tfbkpd}
\eeq
In the S configuration
\beq
\tf_S=-\tf_D
\eeq
and in the DC configuration
\beq
\tf_{DC}=\frac{1}{2}\tf_D.
\eeq

In this case factor 1/2
for the DC contribution is not compensated by the enhanced number
of the interactions between different gluons in the BKP state
found by the cutting plane.
So the final result is $\tf=(1/2)\tf_D$.

\section{Conclusions}
The final inclusive cross-section is obtained from Eqs. (\ref{inclaa})
or (\ref{incldd}) with the high-energy function $F(y,k)$ given by the sum
of all the studied contributions:
\beq
F^{tot}(y,k)=\sum_{i=1}^5F_{i}^P+F^V+F^{2V}+\tilde{F}_{2V}+F^{BKP}+
\tilde{F}^{BKP}.
\label{ftot}
\eeq
Here the particular contributions are:

$F^P{i}$ is emission from pomerons attached to the participants with
$i=1,..5$ corresponding to Figs. \ref{fig1},1,..5, and given by Eqs.
(\ref{fp}), (\ref{fp23}),(\ref{f3p}) and (\ref{fp5}).

$F^V$ is emission from the single interaction in Figs. \ref{fig1},2 and 3,
given by Eq. (\ref{fv})

$F^{2V}$ is emission from one of the two interactions in
Fig. \ref{fig1},4 with the redistribution of colour, Eqs. (\ref{ffh}),
(\ref{ffl}) and (\ref{ffin1})

$\tilde{F}^{2V}$ is emission from one of the two interactions in
Fig. \ref{fig1},5 with the direct transmission of colour, Eqs. (\ref{tffh}),
(\ref{tffl}) and (\ref{bffin1}).

$F^{BKP}$ is emission from the BKP state in
Fig. \ref{fig1},4 with the redistribution of colour, Eqs. (\ref{fbkp})
and (\ref{fbkpd}).

$\tilde{F}^{BKP}$ is emission from the BKP state in
Fig. \ref{fig1},5 with the direct transmission of colour, Eq. (\ref{fbkp})
and  Eq. (\ref{tfbkpd}) divided by two.

Calculation of all these expressions presents a formidable task,
especially including the BKP contributions, which requires solving the
equation for 4 interacting reggeized gluons. However without calculations
it is possible to make some comments about the behaviour of the inclusive
cross-sections at high energies. Note that, unlike the total
cross-sections, the inclusive ones do not involve disconnected parts,
which are cancelltd by the well-know AGK rules.
So the found contributions plus emission from the single pomeron exchange
exhaust the total inclusive cross-sections
in our approximation.

If one assumes that the pomerons attached to the
particpants are indeed given by the standard solutions of the BFKL equation
and grow like $\exp(\Delta_{BFKL} Y)$ with rapidity, then the behaviour of
all terms in the sum (\ref{ftot}) will be determined by the growth of these
attached pomerons, and the inclusive
cross-section will roughly grow as the square of the pomeron,
that is as  $\exp(2\Delta_{BFKL} Y)$, independent of the rapidity $y$
of the observed gluon. In this case appearance of the BKP state will not
significantly change the asymptotical behaviour, since this state grows
slowlier than the pomeron at high energies. However this does not mean that
contribution from the BKP states can be neglected. Rather this state will
enter the cross-section at preasymptotical energies, providing factors
which grow as power of $Y$.

However assumption that the interaction with the particpants is governed
by simple BFKL pomerons is obviously too crude (and violates unitarity).
Substituting simple pomerons by some more realistic expressions,
say, extracted directly
from the experiment or taking into account gluon saturation, will tame the
exponential growth of pomeron legs in Fig. \ref{fig1}. Then the
behaviour of the inclusive cross section will be determined by the
modest growth of the BKP state $\sim \exp{\Delta_{BKP}Y}$ with
$\Delta{BKP}=0.243\Delta_{BFKL},~\cite{KKM}$.
So in this realistic case obsevation of the inclusive gluon production
can provide direct information on the properties of the BKP state
composed of 4 reggeized gluons.

Finally note that this conclusion is truly valid for deuteron-deuteron
scattering. For heavy nuclei scattering the interaction of only two
pairs of nucleons from each nucleus is only a  part of the total
contribution. Generalization of our approach to interactions of any
number of nucleons in each particpant nucleus is postponed for future
studies.

\section{Acknowledgments}

This work has been supported by the RFFI grant 12-02-00356-a
and the SPbSU grants 11.059.2010, 11.38.31.2011 and 11.38.660.2013.


\end{document}